\documentclass[pra,twocolumsn,aps,superscriptaddress]{revtex4-1}
\usepackage{bm}%
\expandafter\ifx\csname package@font\endcsname\relax\else
 \expandafter\expandafter
 \expandafter\usepackage
 \expandafter\expandafter
 
 \expandafter{\csname package@font\endcsname}
\fi

\usepackage{array}
\usepackage{amsmath,amssymb}
\usepackage{MnSymbol}
 \usepackage{float}
\usepackage{graphicx}
\usepackage{lipsum}
\usepackage{graphicx}%
\usepackage{subfigure}
\usepackage{mathrsfs}
\usepackage{bm}
\usepackage{mathtools}
\usepackage{epstopdf}
\usepackage{hyperref}

\hypersetup{%
   pdfpagemode=None, 
   pdfstartpage=1,
   pdfmenubar=true,
   pdftoolbar=true,
   colorlinks = true,
   linkcolor=blue,
   citecolor=blue,
   urlcolor=blue,
   bookmarksopen=false
 }

\usepackage[normalem]{ulem}   
\makeindex

\begin{document}
\title{Polaritons in an electron gas -- quasiparticles and Landau effective interactions}
\author{Miguel Angel Bastarrachea-Magnani}
\affiliation{Departamento de F\'isica, Universidad Aut\'onoma Metropolitana-Iztapalapa, San Rafael Atlixco 186, C.P. 09340, CDMX, M\'exico.}
\affiliation{Department of Physics and Astronomy, Aarhus University, Ny Munkegade, DK-8000 Aarhus C, Denmark}
\author{Jannie Thomsen}
\affiliation{Department of Physics and Astronomy, Aarhus University, Ny Munkegade, DK-8000 Aarhus C, Denmark}
\author{ Arturo Camacho-Guardian}
\affiliation{ T.C.M. Group, Cavendish Laboratory, University of Cambridge, JJ Thomson Avenue, Cambridge, CB3 0HE, U.K.}
\author{Georg M. Bruun}
\affiliation{Department of Physics and Astronomy, Aarhus University, Ny Munkegade, DK-8000 Aarhus C, Denmark}
\affiliation{Shenzhen Institute for Quantum Science and Engineering and Department of Physics, Southern University of Science and Technology, Shenzhen 518055}

\begin{abstract}
Two-dimensional semiconductors inside optical microcavities have emerged as a versatile platform to explore new hybrid light-matter quantum states. 
The strong light-matter coupling leads to the formation of exciton-polaritons, 
which in turn interact with the surrounding electron gas to form  quasiparticles called polaron-polaritons. Here, we develop a general microscopic  framework to calculate 
the properties of these  quasiparticles such as their energy and the interactions between them. From this, we give  
microscopic expressions for the parameters entering a Landau theory for the polaron-polaritons, which  
offers a simple yet powerful way to describe such interacting light-matter many-body systems. As an example of the application of our framework, we then use the ladder approximation to explore the properties of the polaron-polaritons.   We furthermore show that they  can be measured in a non-demolition way via the light transmission/reflection 
spectrum of the system. Finally, we demonstrate that the Landau effective interaction  mediated by   electron-hole excitations 
 is attractive leading to red shifts of  the polaron-polaritons.
Our work provides a systematic framework to study exciton-polaritons in  electronically doped two-dimensional materials such as 
novel  van der Waals heterostructures.
\end{abstract}

%
\maketitle
%
%
%
%
%
\section{Introduction}
%
Semiconductors in optical microcavities constitute a rich setting for exploring hybrid light-matter quantum systems with potential optoelectronic applications~\cite{Sanvitto2016,Kavokin2017}. An important example is the case of exciton-polaritons, which are quantum mechanical superpositions of 
 photons and bound electron-hole pairs confined in a two-dimensional (2D) semiconductor layer inside an optical cavity~\cite{Hopfield1958,Weisbuch1992}. An appealing 
 feature of polaritons is that  they inherit the properties of both their fundamental constituents thereby providing a tunable way to transfer attributes from matter to light, and viceversa. 
Hence, not only can they  be selectively excited, controlled and detected by optical means, but  they also possess strong interactions that introduce novel non-linear optical effects~\cite{Laussy2012,Carusotto2013}. 
As exciton-polaritons can be considered bosons for extended temperature and density ranges , they  exhibit effects like Bose-Einstein condensation and superfluidity~\cite{Kasprzak2006,Balili2007,Amo2009,Amo2009b,Kohnle2011,Kohnle2012,Lagoudakis2008,Sanvitto2010}, although 
the pump-loss nature of the experiments leads to a number of important differences compared to the equilibrium  condensates. 

Atomically thin transition-metal dichalcogenids (TMDs)~\cite{Radisavljevic2011,Mak2016,Wang2018} are among the 2D materials that have been in the spotlight in recent years. They are composed by two hexagonal planes of a transition metal atom M (Mo, W) that covalently binds with chalcogen atoms (S, Se, Te) to form an hexagonal lattice with a trigonal prismatic arrangement ($\mbox{MX}_2$)~\cite{Bromley1972,Mak2010,Zhu2011}. It has been found that
atomically thin layers of TMDs are thermodynamically stable and that they are
direct-gap semiconductors from the visible to the infrarred~\cite{Novoselov2005,Wang2018,Mak2010,Splendiani2010}. The extrema of the bands are located at the finite momentum $K^+$ ($K^-$) points in the hexagonal Brillouin zone and connected by a broken inversion symmetry. Together with a strong spin-orbit coupling (SOC) this leads to valley-spin locking, i.e., the coupling between the valley and spin degrees of freedom~\cite{Ramasubramaniam2012,Xiao2012,Echeverry2016}. As a result, there are  valley selective   optical rules~\cite{Cao2012,Yu2015,Wang2018}, which, together with strong light-matter coupling~\cite{Dufferwiel2015,Liu2015} offer a promising playground for spin optoelectronics and valleytronics~\cite{Xiao2012,Xu2014,Schaibley2016}.

The large binding energy of excitons in TMDs as compared to other microcavity semiconductors like quantum-wells~\cite{Chichibu1996,Qiu2013,He2014}, combined with the  
possibility to control the electron density in the different valleys,
opens up exciting new venues to explore Bose-Fermi mixtures in a hybrid light-matter setting~\cite{Mak2012,Emmanuele2020,Julku2021}.  
This has stimulated a number of studies regarding the properties electron-exciton mixtures and their coupling to light~\cite{Suris2001,Suris2003,Rapaport2001,Qarry2003,Bajoni2006,Pimenov2017,Efimkin2018,Glazov2020,Kyriienko2020,Rana2020,Efimkin2020arXiv}. In particular,
 the emergence of new quasiparticles, the so-called Fermi-polaron-polaritons have been observed~\cite{Sidler2016}. They  
  can be roughly described as a coherent superposition of photons and Fermi polarons, which are formed by the polaritons interacting with the 
surrounding electron gas (2DEG) in analogy with what is observed  in atomic 
gases~\cite{Schirotzek2009,Kohstall:2012tf,Koschorreck12,Cetina2015, Cetina2016,Scazza2017,Adlong2020,Fritsche2021}.  

Two recent experiments have observed large energy  shifts of these polaron-polaritons due to the injection of itinerant electrons in a monolayer TMD 
indicating the presence of induced interactions between them~\cite{Tan2020,Emmanuele2020}, which  opens the door to exploring interacting 
quasiparticles in a new hybrid light-matter setting. 
Landau's theory of quasiparticles stands out as a powerful yet simple framework to  describe precisely such 
interacting many-body systems, including their  single particle and collective properties both in- and out-of-equilibrium~\cite{Landau1957,Landau1957b,Baym1991}.
In light of this, an important question concerns how to calculate  the parameters entering such a Landau theory for polaron-polaritons.

Inspired by this,  we present here a theoretical framework for  polaron-polaritons in a 2DEG in terms of Green's functions.  We moreover show how this can be 
used to calculate the parameters of a Landau theory of polaron-polaritons,  which encompasses the strong light-matter coupling. 
Apart from assuming that the concentration 
of the polaron-polaritons is much smaller than that of the 2DEG and that equilibrium theory can be applied,  our theory is completely general.
We then give a concrete example of  these results by employing an approximate many-body theory, the so-called ladder approximation, which includes  
 strong two-body correlations leading to a bound state between an exciton and an electron, i.e.\ a trimer. Using this, we explore the different polaron-polariton branches and demonstrate how the 
 transmission/reflection spectrum of the system offer a  new experimental way to determine the  energy and residue of the underlying polarons in a non-demolition way.  
The energy of the polaron-polaritons is then shown to decrease with their concentration corresponding to an attractive Landau quasiparticle interaction mediated 
by particle-hole excitations in the 2DEG.



The remainder of the manuscript is structured as follows. In Section.~\ref{sec:2}, we introduce the system and discuss 
the formation of the hybrid light-matter polaritons.   In Section.~\ref{sec:3}, we turn our attention to the effects of interactions and show how this can be described microscopically. We then connect this to Landau's quasiparticle theory providing microscopic expressions for the quasiparticle energies and their effective 
interactions.  In Sec.~\ref{sec:4}, we apply these results to the ladder approximation, and analyse the predicted properties 
of the quasiparticles and the interactions between them. We also propose a new way to measure those  via the light transmission/reflection spectrum. Finally, in Sec.~\ref{sec:5} we present our conclusions and offer some perspectives.



\section{System}
\label{sec:2}
We consider a  2D semiconductor in an optical microcavity. Photons in the cavity are strongly coupled to  excitons in the semiconductor and the excitons in turn interact with a 2D electron gas (2DEG). The Hamiltonian for the system is  $\hat H=\hat H_0+\hat H_I$ where  
\begin{align}
\hat{H}_{0}=\sum_{\mathbf{k}}\left[\varepsilon_{e\mathbf{k}}\hat e_{\mathbf{k}}^\dagger\hat e_{\mathbf{k}}+\varepsilon_{x\mathbf{k}}\hat{x}_{\mathbf{k}}^{\dagger}\hat{x}_{\mathbf{k}}+\varepsilon_{c\mathbf{k}}\hat{c}_{\mathbf{k}}^{\dagger}\hat{c}_{\mathbf{k}}\right]+ \sum_{\mathbf{k}}\Omega\left(\hat{x}_{\mathbf{k}}^{\dagger}\hat{c}_{\mathbf{k}}+\hat{c}_{\mathbf{k}}^{\dagger}\hat{x}_{\mathbf{k}}\right)
\label{eq:Hlm}
\end{align} 
{are the non-interacting and the light-matter coupling terms}. Here  $\hat{x}_{\mathbf{k}}^{\dagger}$, $\hat{c}_{\mathbf{k}}^{\dagger},$ and  $\hat{e}_{\mathbf{k}}^{\dagger},$ creates an exciton,  photon, and electron respectively with two-dimensional crystal momentum $\mathbf{k}$. The energy of these particles is  $\varepsilon_{x\mathbf{k}}=\mathbf{k}^{2}/2m_{x},$ $\varepsilon_{c\mathbf{k}}=\mathbf{k}^{2}/2m_{c}+\delta$, and  $\varepsilon_{e\mathbf{k}}=\mathbf{k}^{2}/2m_{e}$, where  $m_x$, $m_c$,  and $m_e$ is their mass and  $\delta$ is the detuning between the exciton and photon energies at zero momentum. We set $\hbar=k_B=1$ throughout. For concreteness, we take $m_{c}=10^{-5}m_{x}$, $m_{x}=2m_{e}$ and assume the light-matter coupling $\Omega$ to be real. The energy offset of the electrons will be absorbed into their chemical potential. It follows from the  optical and valley selection rules of TMDs~\cite{Radisavljevic2011,Mak2016,Wang2018} that polarised photons couple to excitons in a specific spin  and valley state, which in turn predominantly interacts with the 2DEG in the opposite valley. Here, we focus on a given spin and valley and therefore suppress those degrees of freedom in Eq.~\eqref{eq:Hlm} and onwards. The excitons are assumed to have a binding energy much larger than any other relevant energy scale in the system so that they can be considered as point bosons. For high exciton densities or localised excitons, their composite nature becomes important and the point boson approximation breaks down, leading to changes in the effective light-matter interaction and saturation effects~\cite{Emmanuele2020,Kyriienko2020}.

The non-interacting Hamiltonian Eq.~\eqref{eq:Hlm} is readily diagonalised by means of a Hopfield transformation~\cite{Hopfield1958} 
\begin{align}
\begin{bmatrix}
\hat{x}_{\mathbf{k}} \\
\hat{c}_{\mathbf{k}}
\end{bmatrix}
=
\begin{bmatrix}
\mathcal{C}_{\mathbf{k}} & -\mathcal{S}_{\mathbf{k}} \\
\mathcal{S}_{\mathbf{k}} & \mathcal{C}_{\mathbf{k}} \\
\end{bmatrix}
\begin{bmatrix}
\hat{L}_{\mathbf{k}} \\
\hat{U}_{\mathbf{k}}
\end{bmatrix}
\label{eq:Polaritons}
\end{align}
where $L_{\mathbf{k}}^{\dagger}$ ($U_{\mathbf{k}}^{\dagger}$) are the creation operators of lower and upper polaritons respectively with momentum $\mathbf k$. The corresponding Hopfield coefficients are $\mathcal{C}_{\mathbf{k}}^{2}=(1+\delta_{\mathbf{k}}/\sqrt{\delta_{\mathbf{k}}^{2}+4\Omega^{2}})/2$ and
$\mathcal{S}_{\mathbf{k}}^{2}=1-\mathcal{C}_{\mathbf{k}}^{2}$ with $\delta_{\mathbf{k}}=\varepsilon_{c\mathbf{k}}-\varepsilon_{x\mathbf{k}}$, and  
\begin{align} \label{eq:Polaritonenergies}
\varepsilon_{\sigma\mathbf{k}}=\frac12\left(\varepsilon_{c\mathbf{k}}+\varepsilon_{x\mathbf{k}}\pm\sqrt{\delta_{\mathbf{k}}^{2}+4\Omega^{2}}\right),
\end{align}
giving the energy of the standard upper $\sigma=\text{U}$ and lower $\sigma=\text{L}$ exciton-polaritons in absence of the Fermi sea. Interactions between the excitons and electrons in opposite valleys are described by the term 
 \begin{align}
\hat{H}_I={\frac{1}{\mathcal A}}\sum_{\mathbf{q},\mathbf{k},\mathbf{k}'}V_{\mathbf{q}}\hat{e}_{\mathbf{k}+\mathbf{q}}^{\dagger}\hat{x}_{\mathbf{k}'-\mathbf{q}}^{\dagger}
\hat{x}_{\mathbf{k}'}\hat{e}_{\mathbf{k}},
\label{eq:Hamiltonian}
\end{align} 
where $\mathcal A$ is the area of the system. 
For small Fermi energies and relevant momenta the electron-exciton interaction can be approximated as a contact one $V_{\mathbf{q}}\simeq \mathcal{T}_{0}$~\cite{Sidler2016}. This is equivalent to treating the exciton-polaritons as point-like bosons. Also, we assume that the Coulomb interaction between the electrons are included by a renormalisation of their dispersion using Fermi liquid theory~\cite{Shankar1994,ahn2021fragile}, and we furthermore neglect the direct interaction between excitons. For small densities, the latter is rather weak due to the large binding energy of the excitons, which is typically two orders of magnitude larger than the rest of energy scales~\cite{Chichibu1996,Qiu2013,He2014,Zhu2015}, and it can easily be included at the mean-field level. 

\section{Fermi polaron-polaritons}
\label{sec:3}
We now consider the situation where the density of exciton-polaritons is small compared to the electron density. In this case, the effects of the exciton-polaritons on the 2DEG can be neglected and the problem reduces to that of mobile bosonic impurities in an electron gas. The interaction between the exciton-polaritons and the surrounding electron gas then gives rise to the formation of quasiparticles denoted Fermi polaron-polaritons or, in short, polaron-polaritons. Apart from the presence of strong light coupling this has strong similarities to the formation of Fermi polarons in atomic gases~\cite{Massignan2014}. In this section we will describe their generic properties both from a microscopic point of view as well as using Landau's quasiparticle framework. We will furthermore provide precise links between the two descriptions when appropriate. While these results are general, we will illustrate them by using a microscopic approximated many-body theory as an example. 

\subsection{Microscopic theory}
Despite the fact that polariton systems are driven by external lasers, many of their steady-state properties can be accurately described using equilibrium theory with a few modifications, such as chemical potentials being determined by the external laser frequencies~\cite{Carusotto2013}. We therefore employ  finite temperature quantum field theory to analyse the problem microscopically~\cite{Fetter1971}. Since the electrons are unaffected by the excitons, we can focus on the cavity photons and excitons described by 
the $2\times 2$ exciton-photon finite-temperature Green's function ${\mathcal G}(\mathbf k,\tau)=-\langle T_\tau \{\hat\Psi_{\mathbf k}(\tau)\hat\Psi_{\mathbf k}^\dagger(0)\}\rangle$, where $\hat\Psi_{\mathbf k}=[\hat x_{\mathbf k},\hat c_{\mathbf k}]^T$ and $T_\tau$ denotes the imaginary time ordering. By Fourier transforming, it can be written in terms of the free propagator $\mathcal{G}_{0}(k)$ and the proper self-energy $\mathbf{\Sigma}(k)$ as
\begin{align}
{\mathcal G}^{-1}(k)={\mathcal G_{0}}^{-1}(k)-\mathbf{\Sigma}(k)=
\begin{bmatrix}
i\omega_l-\varepsilon_{x\mathbf{k}}& 0 \\
0 & i\omega_l-\varepsilon_{c\mathbf{k}}
\end{bmatrix}
-
\begin{bmatrix}
\Sigma_{xx}(k) & \Omega  \\
\Omega & 0
\end{bmatrix}.
\label{eq:GreensFn}
\end{align} 
where $k=(\mathbf{k},\omega_{l})$, $\omega_l=2\pi lT$ with $l=0,\pm1,\ldots$ is a bosonic Matsubara frequency, $T$ is the temperature, and $\Sigma_{xx}(k)$ is the exciton self-energy. As usual, one can obtain the retarded Green's function by  analytic continuation $\mathcal G(\mathbf{k},\omega)=\left.\mathcal{G}(\mathbf{k},i\omega_{l})\right|_{i\omega_{l}\rightarrow\omega+i0^{+}}$. 

In the absence of light, the problem is equivalent to impurity particles interacting with a Fermi sea, which is known to lead to the formation quasiparticles called Fermi polarons~\cite{Chevy2006,Massignan2014,levinsen2015}. The coupling to light turns these polarons into polaron-polaritons, and in analogy with Eq.~\eqref{eq:Polaritonenergies} the  energy of these quasiparticles is given by the self-consistent solutions of
\begin{align} 
\varepsilon_{{\sigma}\mathbf{k}}=\frac{1}{2}\left[\varepsilon_{c\mathbf{k}}+\varepsilon_{x\mathbf{k}}+\Sigma_{xx}\left(\mathbf{k},\varepsilon_{\sigma\mathbf{k}}\right)\pm
\sqrt{\left[\delta_{\mathbf{k}}-\Sigma_{xx}\left(\mathbf{k},\varepsilon_{\sigma\mathbf{k}}\right)\right]^{2}+4\Omega^{2}}\right].
\label{eq:poles}
\end{align}
Here, the subindex $\sigma$ denotes the different quasiparticle branches emerging in the system. 
Also, a new set of Hopfield coefficients arise giving the matter and photon components of the polaron-polaritons. As in Eq.~\eqref{eq:Polaritons} they are 
\begin{align}
\mathcal{C}_{\mathbf{k}\sigma}^{2}=\frac12+\frac{\varepsilon_{c\mathbf{k}}-\varepsilon_{x\mathbf{k}}-\Sigma_{xx}\left(\mathbf{k},\varepsilon_{\sigma\mathbf{k}}\right)}{2\sqrt{\left[\varepsilon_{c\mathbf{k}}-\varepsilon_{x\mathbf{k}}-\Sigma_{xx}\left(\mathbf{k},\varepsilon_{\sigma\mathbf{k}}\right)\right]^{2}+4\Omega^{2}}}\hspace{0.5cm}\text{ and }\hspace{0.5cm}\mathcal{S}_{\mathbf{k}\sigma}^{2}=1-\mathcal{C}_{\mathbf{k}\sigma}^{2}.
 \label{eq:HopS}
\end{align}

\subsection{Landau theory}\label{sec:Landau}
Landau's description of macroscopic systems in terms of quasiparticles   is  a highlight in theoretical physics and 
provides a remarkably simple yet accurate description of  otherwise complex many-body 
systems~\cite{Landau1957,Landau1957b}. This includes both their single-particle and collective equilibrium and non-equilibrium properties, and it is therefore 
important to understand how it can be applied to polaron-polaritons. 
We now address this question and  provide precise links between Landau's framework and  the microscopic theory in the previous section. 

The foundation of Landau's theory idea is to write the energy $E$ of a system in powers of its low energy excitations, which have particle like properties, i.e.\ the quasiparticles as~\cite{Baym1991}
\begin{align}
\label{LandauT}
E=E_g+\sum_{\mathbf{q},\sigma}\varepsilon_{\mathbf k\sigma}^0 n_{\mathbf k\sigma}+\frac{1}{2{\mathcal A}}\sum_{\mathbf{k},\mathbf{k}',\sigma,\sigma'}
\mathsf f_{\mathbf k\sigma,\mathbf k'\sigma'} n_{\mathbf k\sigma} n_{\mathbf k'\sigma'}+...,
\end{align}
where $E_g$ is the ground state energy of the system and $\varepsilon_{\mathbf{k}\sigma}^{0}$ is the quasiparticle energy.  The  distribution function in a given quasiparticle branch $\sigma$ is given by $ n_{\mathbf k\sigma}$, and $\mathsf f_{\mathbf k\sigma,\mathbf k'\sigma'}$ is the interaction between quasiparticles in branches $\sigma$ and $\sigma'$ with momenta $\mathbf{k}$ and $\mathbf{k}'$. In principle, there are terms of higher order in $n_{\mathbf k\sigma}$ in Eq.~\eqref{LandauT}, which correspond to three-body interaction terms and higher. 
However, such terms are usually not important for realistic densities and it is standard in Landau's quasiparticle theory to truncate the series at quadratic order 
corresponding to including two-body interactions, as we do here. 

In the present case, the quasiparticles are  the polaron-polaritons and their energy $\varepsilon_{\mathbf k\sigma}^0$ are given by solutions of Eq.\ \eqref{eq:poles} taking the zero impurity limit, i.e.\ a vanishing  quasiparticle distribution function $n_{\mathbf k\sigma}=0$. The ground state of the system is simply the 2DEG with no polaron-polaritons present with the energy
 ${\mathcal A}n_{e}\varepsilon_F/2$ where $n_e$ is the density of the 2DEG with Fermi energy $\varepsilon_F$. When the number of quasiparticles is non-zero, it follows from Eq.~\eqref{LandauT} that their energy is 
\begin{align}
\varepsilon_{\mathbf k\sigma}=\varepsilon_{\mathbf k\sigma}^0+\frac1{\mathcal A}\sum_{\mathbf k'\sigma'}\mathsf f_{\mathbf k\sigma,\mathbf k'\sigma'} n_{\mathbf k'\sigma'}.
\label{eq:LandauQPenergy}
\end{align} 
It follows from Eq.~\eqref{eq:LandauQPenergy} that the interaction between the quasiparticles can be found as~\cite{Bastarrachea2021}
\begin{align} \label{eq:LandauInt}
\frac{\mathsf f_{\mathbf{k}\sigma,\mathbf{k}'\sigma'}}{\mathcal A}=\frac{d \varepsilon_{\mathbf k\sigma}}{d n_{\mathbf k'\sigma'}}=
Z_{\mathbf{k}\sigma}\mathcal{X}_{\mathbf{k}\sigma}^{2}\frac{\partial\Sigma_{xx}(\mathbf{k},\varepsilon_{\mathbf{k}\sigma})}{\partial n_{{\mathbf{k}}'\sigma'}},
\end{align}
where 
\begin{gather}Z_{\mathbf{k}\sigma}^{-1}=1-\mathcal{X}_{\mathbf{k}\sigma}^{2}\partial_\omega\Sigma_{xx}(\mathbf{k},\varepsilon_{\mathbf{k}\sigma})\end{gather} is the residue of a polaron-polariton in 
branch $\sigma$ with momentum $\mathbf k$ and we have used  Eq.~\eqref{eq:poles} in the second equality. 
Here, $\mathcal{X}_{\mathbf{k}\sigma}=\mathcal{S}_{\mathbf{k}\sigma}$ when the quasiparticle energy is determined  using the $+\sqrt{\ldots}$  version of the upper polariton poles in Eq.~\eqref{eq:poles}, 
whereas $\mathcal{X}_{\mathbf{k}\sigma}=\mathcal{C}_{\mathbf{k}\sigma}$ when the $-\sqrt{\ldots}$  version of the lower polariton in Eq.~\eqref{eq:poles} is used.  Compared to the 
usual microscopic many-body formula for  Landau's quasiparticle interaction~\cite{Camacho2018,Giuliani2005},  Eq.\ \eqref{eq:LandauInt} has the additional feature of 
containing the many-body Hopfield coefficients. They reflect that it is only the excitonic part of the quasiparticles which interact with the surrounding 2DEG. 

Equations \eqref{eq:GreensFn}-\eqref{eq:LandauInt} provide a framework for describing polaron-polaritons in a 2DEG microscopically and moreover 
show how to connect this to Landau's quasiparticle theory. The main assumptions are that the concentration of polaron-polaritons is much smaller than that of the electrons so that their effects on the 2DEG can be neglected, and that we can use equilibrium theory to describe its steady state properties. We now illustrate these results using an approximate many-body theory.  

\section{The ladder approximation}
\label{sec:4}
To give a concrete example of the results in the previous section, we apply the much used so-called ladder approximation to describe polaritons interacting with a 2DEG. This theory has turned out to be surprisingly accurate for mobile impurities in atomic Fermi gases~\cite{Massignan2014}, which is a problem with many similarities to the one at hand. The basic idea is to include the two-body scattering physics exactly in a many-body environment  and it is thus particularly suited to describe systems with strong two-body correlations such as molecule formation or hard core repulsion~\cite{Fetter1971}. In the present context, the molecules correspond to bound states of an exciton and an electron, i.e.\ a trion,  which indeed have been observed in TMDs~\cite{Lampert1958,Thilagam1997,Esser2001,Mak2012,Courtade2017,Zhu2014,Ganchev2015,Sie2016} motivating the use of this approximation. In the ladder approximation, the exciton self-energy is given by 
\begin{eqnarray} \label{eq:4}
\Sigma_{xx}(k)=\frac{T}{\mathcal{A}}\sum_{q}\mathcal{G}_{e}(q)\mathcal{T}(k+q),
\end{eqnarray}
where  $k=(\mathbf k,i\omega_l)$, $\mathcal{G}^{-1}_{e}(\mathbf{k},i\omega_{j})=i\omega_{j}-\xi_{\mathbf{k}}^{e}$ is the electron propagator
with $i\omega_{j}=(2j+1)\pi T$  a fermionic Matsubara frequency, and $\sum_q$ denotes a sum over both Matsubara frequencies and 2D momentum. The electron  energy is taken with respect to the Fermi energy of the 2DEG, i.e., $\xi_{e\mathbf{k}}=\varepsilon_{e\mathbf{k}}-\varepsilon_{F}$. In Eq.~\eqref{eq:4}, we have introduced the  exciton-electron scattering matrix  given by~\cite{Wouters2007,Carusotto2010,Bastarrachea-Magnani2019}
\begin{align} 
\mathcal{T}(k)
=\frac1{\text{Re}\Pi_V(\mathbf k=0,\varepsilon_{T})-\Pi(k)},
\label{eq:Tmatrix}
\end{align} 
where  $\Pi(k)$ is the in-medium exciton-electron pair-propagator
\begin{align} \label{eq:5}
\Pi(k)=-\frac{T}{\mathcal{A}}\sum_{q}\mathcal{G}^{(0)}_{xx}(k+q)\mathcal{G}_{e}(-q)=
\sum_{\sigma}\int\! \frac{d^{2}\mathbf{q}}{(2\pi)^{2}}\mathcal{X}^{2}_{\sigma\mathbf{k}+\mathbf{q}}\frac{1+n_{B}(\xi_{\mathbf{k}+\mathbf{q}\sigma})-n_{F}(\xi_{e-\mathbf{q}}) }{i\omega_{j}-\xi_{\mathbf{k}+\mathbf{q}\sigma}-\xi^{e}_{-\mathbf{q}}}.
\end{align}
Here, $\mathcal{G}^{(0)}_{xx}(k)=\sum_{\sigma}\mathcal{X}^{2}_{\sigma\mathbf{k}}/(i\omega_l-\xi_{\mathbf k\sigma})$ is the exciton Green's function in the absence of interactions expressed in terms of the upper $\sigma=U$ and lower polariton $\sigma=L$ with $\xi_{\mathbf k\sigma}=\varepsilon_{\mathbf{k}\sigma}-\mu_\sigma$ where $\varepsilon_{\mathbf{k}\sigma}$ is given by Eq.~\eqref{eq:Polaritonenergies}. 
In this way, we  include the hybridisation of the exciton and the photon in the scattering matrix. Note that we have introduced the chemical potentials $\mu_\sigma$ to account for a non-zero concentration of the polaritons described by the Bose-Einstein distribution $n_{B}(x)=[\exp(\beta x)-1]^{-1}$, whereas $n_{F}(x)=[\exp(\beta x)+1]^{-1}$ is the Fermi-Dirac distribution for the electrons. 

In deriving Eq.~\eqref{eq:Tmatrix} we have assumed a momentum independent exciton-electron interaction, which is accurate for  $k_{F}a_{B}^{x}\ll1$, where $a_{B}^{x}$ is the Bohr radius giving the typical size of the exciton. Also, the bare coupling strength has been expressed in terms of the  energy $\varepsilon_{T}$ of the  trion in the absence of the 2DEG as $\text{Re}\Pi_V(0,\varepsilon_T)=\mathcal{T}_{0}^{-1}$~\cite{Wouters2007,Carusotto2010,Bastarrachea-Magnani2019}.
At the level of a single impurity and zero temperature, the $\mathcal T$-matrix formalism is equivalent to Chevy's variational ansatz~\cite{Chevy2006}, which has recently been employed to explore Fermi polaron-polaritons in TMD monolayers~\cite{Sidler2016}. As we shall demonstrate below, our field-theoretical  approach is however readily extended to include the effects of temperature and a non-zero quasiparticle concentration. Such effects are usually challenging to incorporate in a variational approach.
  
\subsection{Zero polaron-polariton density}
\begin{figure}[!ht]
\begin{center}
\begin{tabular}{c}
\includegraphics[width=0.9\columnwidth]{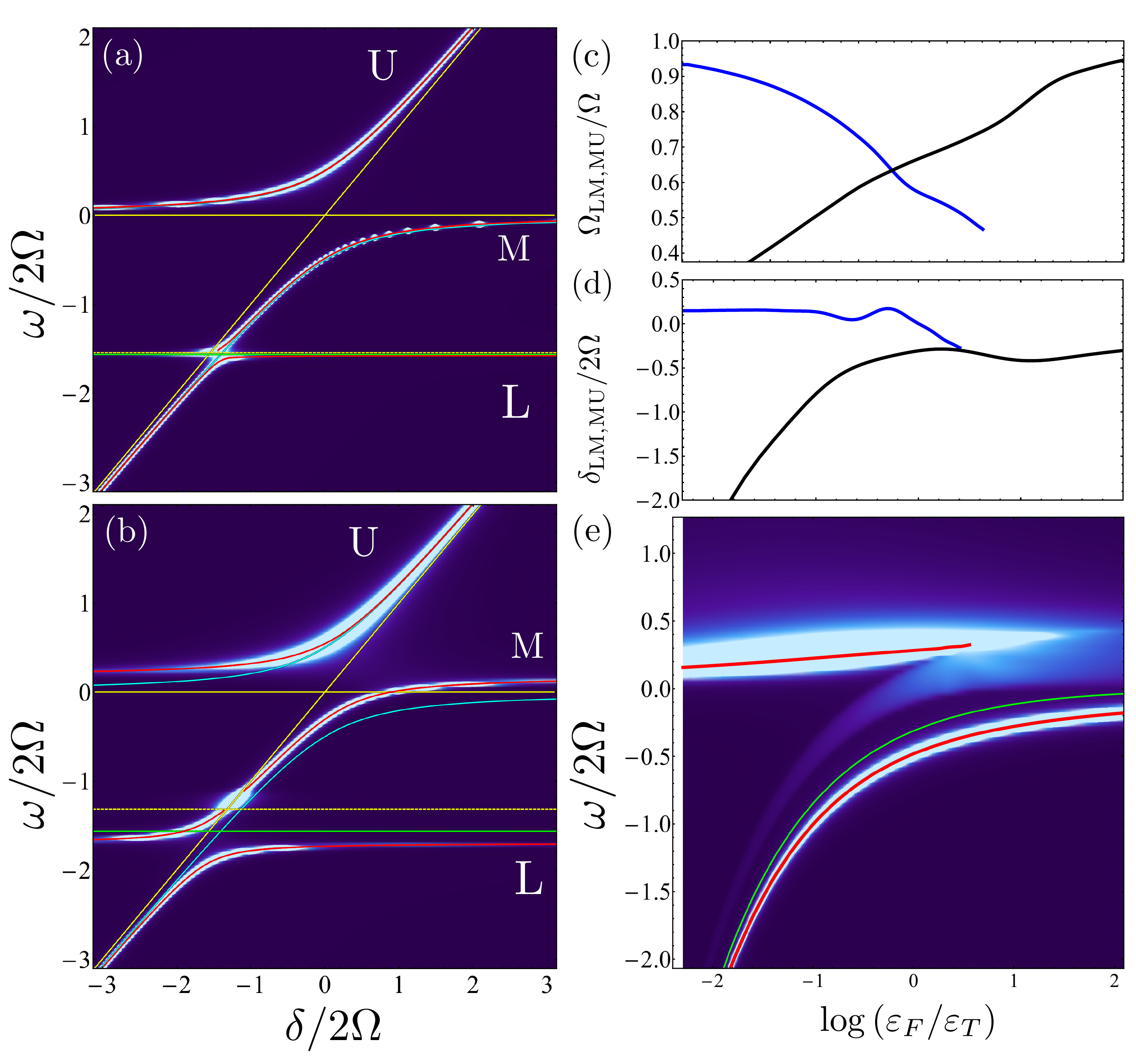} 
\end{tabular}
\end{center}
\vspace{-20pt}
\caption{Photon spectral distribution $A_{cc}(\mathbf{k}=0,\omega)$ for $n_{e}=8.0\mbox{x}10^{10}$ ($\varepsilon_F/\varepsilon_T=0.015$) (a) and $1.0\mbox{x}10^{12}$ ($\varepsilon_F/\varepsilon_T=0.19$) (b). We  observe three quasiparticle branches $\text{L}$, $\text{M}$ and $\text{U}$ of  exciton-polaron-polaritons (red curves). The yellow solid curves correspond to the uncoupled photon and exciton energies, while the cyan lines give the polariton branches in absence of electron-exciton interactions. The horizontal green solid line indicates the bare binding energy of the trion $\varepsilon_{T}$ and the dashed yellow  the binding energy in the presence of many-body correlations. (c) Size of the Rabi coupling for the $\text{L-M}$ branches (attractive polaron) $\Omega_{\text{LM}}$ (blue) and the $\text{M-U}$ branches (repulsive polaron) $\Omega_{\text{MU}}$ (black) as a function of the ratio $\varepsilon_{F}/\varepsilon_{T}$.
(d) Value of the detuning where the avoided crossings between the polaron-polariton branches   occur with the same color coding as in (c). The background colors show the 2D polaron spectral function in the absence of light. For the calculations we employ an additional artificial broadening $\eta/2\Omega=0.01$.}
\label{fig:1} 
\end{figure}

We now discuss the properties of polaron-polaritons in  the limit where their density vanishes, which corresponds to taking $n_{B}(\xi_{\sigma\mathbf{k}+\mathbf{q}})\rightarrow 0$ in Eq.~\eqref{eq:5}. In this case, the Matsubara sum in Eq.~\eqref{eq:4} yields 
\begin{align} \label{eq:ss0}
\Sigma_{xx}(k)= \int\!\frac{d^{2}\mathbf{q}}{(2\pi)^{2}}n_{F}(\xi_{e\mathbf{q}})\mathcal{T}(\mathbf{k}+\mathbf{q},i\omega_{\nu}+\xi_{e\mathbf{q}}).
\end{align}
In Fig.~\ref{fig:1}, we show the zero momentum photonic spectral density $A_{cc}(\omega)=-2\mbox{Im}G_{cc}(\mathbf{k}=0,\omega)$ as a function of the detuning $\delta$ obtained by inverting  Eq.~\eqref{eq:GreensFn}. We use the experimentally  realistic values  $\Omega=8\mbox{meV}$ and $\varepsilon_{T}=-25\mbox{meV}$~\cite{Mak2012,Wang2015}. In Fig.~\ref{fig:1} (a)-(b) we show the spectral function for increasing electron densities with $\varepsilon_F/\varepsilon_T=0.015$ ($n_e=8.0\mbox{x}10^{10}$) and 0.19 ($n_e=1.0\mbox{x}10^{12}$), respectively.  For a typical experimental temperature $T\approx 1K,$~\cite{Tan2020} the thermal energy  remains much smaller than the Rabi coupling $(k_BT/\Omega\approx 0.05)$,  the trion binding energy, and the  Fermi energy of the system. Temperature effects are therefore expected to be negligible. 

Let us first focus on the limit $\delta\gg |\Omega|$ where the photon is decoupled from the excitons and electrons. In addition to the photon, there are two quasiparticle branches in this limit: The so-called attractive polaron corresponding to the exciton attracting the electrons around it giving a quasiparticle energy below the trion energy, and the repulsive polaron corresponding to the electron repelling the electrons around it giving an energy above zero. We see that the repulsive polaron has most spectral weight for low electron density with $\varepsilon_F/\varepsilon_T=0.015$, whereas the attractive branch starts to gain most spectral weight for high electron density with $\varepsilon_F/\varepsilon_T=0.19$. This is consistent with what it is found for polarons in atomic gases, since a small electron density with $\varepsilon_F\ll \varepsilon_T$ corresponds to the so-called BEC limit and a large electron density $\varepsilon_F\gg \varepsilon_T$ corresponds to the BCS limit. For atomic gases, one indeed has that the residue of the attractive polaron approaches unity in the BCS limit whereas that of the repulsive polaron vanishes and vice versa in the BEC limit~\cite{Schmidt2012,Massignan2014}.

\begin{figure}[!ht]
\begin{center}
\begin{tabular}{c}
\includegraphics[width=0.4\columnwidth]{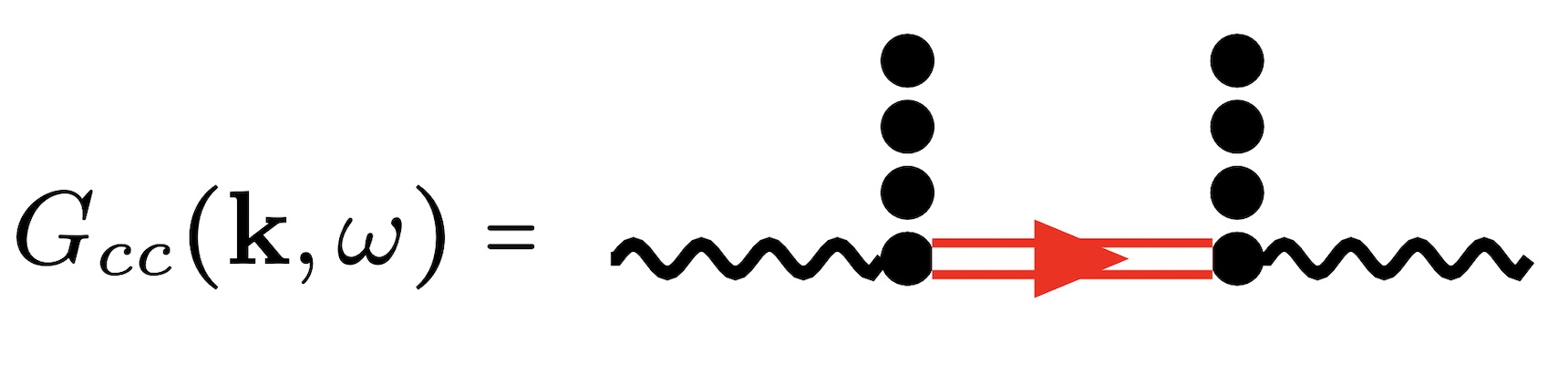} 
\end{tabular}
\end{center}
\vspace{-20pt}
\caption{Feynman diagram for the coupling of the  photon propagator (black, wavy line) to the exciton (red  line). The dotted lines represent the Rabi coupling. }
\label{fig:2} 
\end{figure}

When $\delta/|\Omega|$ decreases, the photon starts to couple to the attractive and repulsive polarons resulting in  three hybrid light-matter quasiparticle branches, which we have denoted as the upper U, middle M, and lower L polaron-polaritons. There are two prominent avoided crossings between these branches as it can be seen in Fig.~\ref{fig:1} (a)-(b), and their size and position can be understood as follows. In absence of any light-matter coupling, the impurity forms an attractive (repulsive) polaron with energy $\omega^{a(r)}_{\mathbf k}$ and residue $Z^{a(r)}_{\mathbf{k}}$~\cite{Schmidt2012,Massignan2014,Efimkin2018}. The coupling of these polarons to the photon can be described by the photon Green's function 
\begin{align}
G^{-1}_{cc}(\mathbf k,\omega)\approx\omega-\varepsilon_{c\mathbf k}-\Omega^2\left[\frac{Z^a_{\mathbf{k}}}{\omega-\omega^a_{\mathbf k}}+\frac{Z^r_{\mathbf{k}}}{\omega-\omega^r_{\mathbf k}}\right], 
\label{Gccpole}
\end{align}
which is illustrated in Fig.~\ref{fig:2}. It describes the repeated transitions between the photon and the polarons by the Rabi coupling as the polaron-polariton propagates through the medium. 
Equation \eqref{Gccpole}  includes only the quasiparticle peaks of the exciton propagator and  ignores any many-body continuum of states in the spirit of Landau theory. From Eq.~\eqref{Gccpole}, we see that the matrix element giving the size of the avoided crossing of the photon branch with the repulsive and attractive polarons is 
\begin{align}
\Omega_{\text{UM}}=\Omega\sqrt{Z^{r}_{\mathbf{k}}}\hspace{1cm}\text{ and }\hspace{1cm}\Omega_{\text{LM}}=\Omega\sqrt{Z^{a}_{\mathbf{k}}},
\label{eq:avoided}
\end{align}
respectively. This explains why the avoided crossing for the repulsive/attractive polaron is large/small for small electron density $\varepsilon_F/\varepsilon_T=0.015$ in  Fig.~\ref{fig:1}(a), since this corresponds to the BEC limit where the residue of the repulsive polaron approaches unity~\cite{Schmidt2012,Massignan2014}. In the same fashion, the avoided crossing of the repulsive/attractive polaron is small/large for large electron density in Fig.~\ref{fig:1}(b), since this corresponds to the BCS limit where the attractive polaron has a residue close to unity and the residue of the repulsive polaron vanishes. 

To explore this further, we plot in Fig.~\ref{fig:1}(c) the size of the two avoided crossings extracted as the minimum energy difference between the polaron-polariton branches 
as a function of the electron density. This clearly shows how $\Omega_{\text{UM}}$ decreases with increasing electron density reflecting the decreasing weight of the repulsive polaron. As the BCS limit is approached, the repulsive polaron becomes ill-defined and we cannot determine  $\Omega_{\text{UM}}$. Mirroring this, $\Omega_{\text{LM}}$ increases with increasing electron density since the residue of the attractive polaron increases as the BCS limit is approached. Since the avoided crossing of the photon with the exciton in the absence of electrons is given by $\Omega$, we conclude from this that the residues of the repulsive and attractive polarons can be extracted by measuring the size of their avoided crossings. 

Furthermore, from Eq.~\eqref{eq:avoided} we see that the position of the avoided crossings is determined by when the energies of the attractive and repulsive polarons cross the photon branch.  
To illustrate this, we plot in Fig.~\ref{fig:1}(d) the value of the detuning where the avoided crossings occur as a function of the electron density. We also plot the spectral function of the 
polaron in a 2D Fermi gas in the absence of light coupling  determined from Eq.~\eqref{eq:GreensFn} setting $\Omega=0$~\cite{Schmidt2012}. The {good} agreement between the 
peaks of this spectral function giving the energies of the attractive and repulsive polarons in a Fermi gas and the positions of the two avoided crossings confirms 
that the underlying physics indeed is driven by the coupling of polarons to light. 

In conclusion, these results  unfold a new experimental way to determine the energy and residue of the polaron in a non-demolition way by detecting the light transmission/reflection spectrum of the system. 
This method represents  an important alternative to earlier approaches based on Rabi-oscillations in radio-frequency (RF) spectroscopy~\cite{Nascimbene2009,Kohstall:2012tf,Scazza2017,Camacho2020a}.
We note that these avoided crossings  have already been observed experimentally~\cite{Sidler2016,Efimkin2017,Tan2020}. 


\subsection{Non-zero polaron-polariton density}
We now consider the case of a non-zero polaron-polariton density focusing on how this affects their energy. From this, we will derive a microscopic expression for Landau's quasiparticle interaction within the ladder approximation. 

Our starting point is Eq.~\eqref{eq:4} for the exciton self-energy. For a non-zero density of excitons, evaluating the Matsubara sum yields~\cite{Bastarrachea2021} 
\begin{align}
\Sigma_{xx}(\mathbf{k},i\omega_{\nu})=& \int\!\frac{d^{2}\mathbf{q}}{(2\pi)^{2}}\Big[n_{F}(\xi_{e\mathbf{q}})\mathcal{T}(\mathbf{k}+\mathbf{q},i\omega_{\nu}+\xi_{e\mathbf{q}})
\nonumber  \\ 
&+
\int_{-\infty}^{\infty}\!\frac{d\omega'}{\pi}\,\frac{n_{F}(\omega')\mbox{Im}\mathcal{T}(\mathbf{k}+\mathbf{q},\omega'+i0^{+})}{i\omega_{\nu}-\omega'+\xi_{e\mathbf{q}}}
-
\frac{n_{F}(\omega^\text{tr}_{\mathbf{k}+\mathbf{q}})Z^\text{tr}_{\mathbf{k}+\mathbf{q}}}{i\omega_{\nu}-\omega^\text{tr}_{\mathbf{k}+\mathbf{q}}+\xi_{e\mathbf{q}}}
\Big].
 \label{eq:ss}
\end{align}
Compared to Eq.~\eqref{eq:ss0}, the finite exciton density gives rise to the two new terms in the second line of Eq.~\eqref{eq:ss}. The last term is a contribution coming from a non-zero population of the trion state, which appears as a pole in the many-body scattering matrix at the energy $\omega^\text{tr}_\mathbf{k}$ with residue $Z^\text{tr}_{\mathbf{k}}$. This results in an interaction between the trions and the excitons mediated by the exchange of an electron~\cite{Bastarrachea2021}, which has been observed to give rise to large optical non-linearities. We neglect this term in the following assuming a zero population of trions and refer the reader to Ref.~\cite{Emmanuele2020} for an analysis of the interesting interaction between excitons and trions mediated by electron exchange. 

\begin{figure}[!ht]
\begin{center}
\begin{tabular}{c}
\includegraphics[width=0.7\columnwidth]{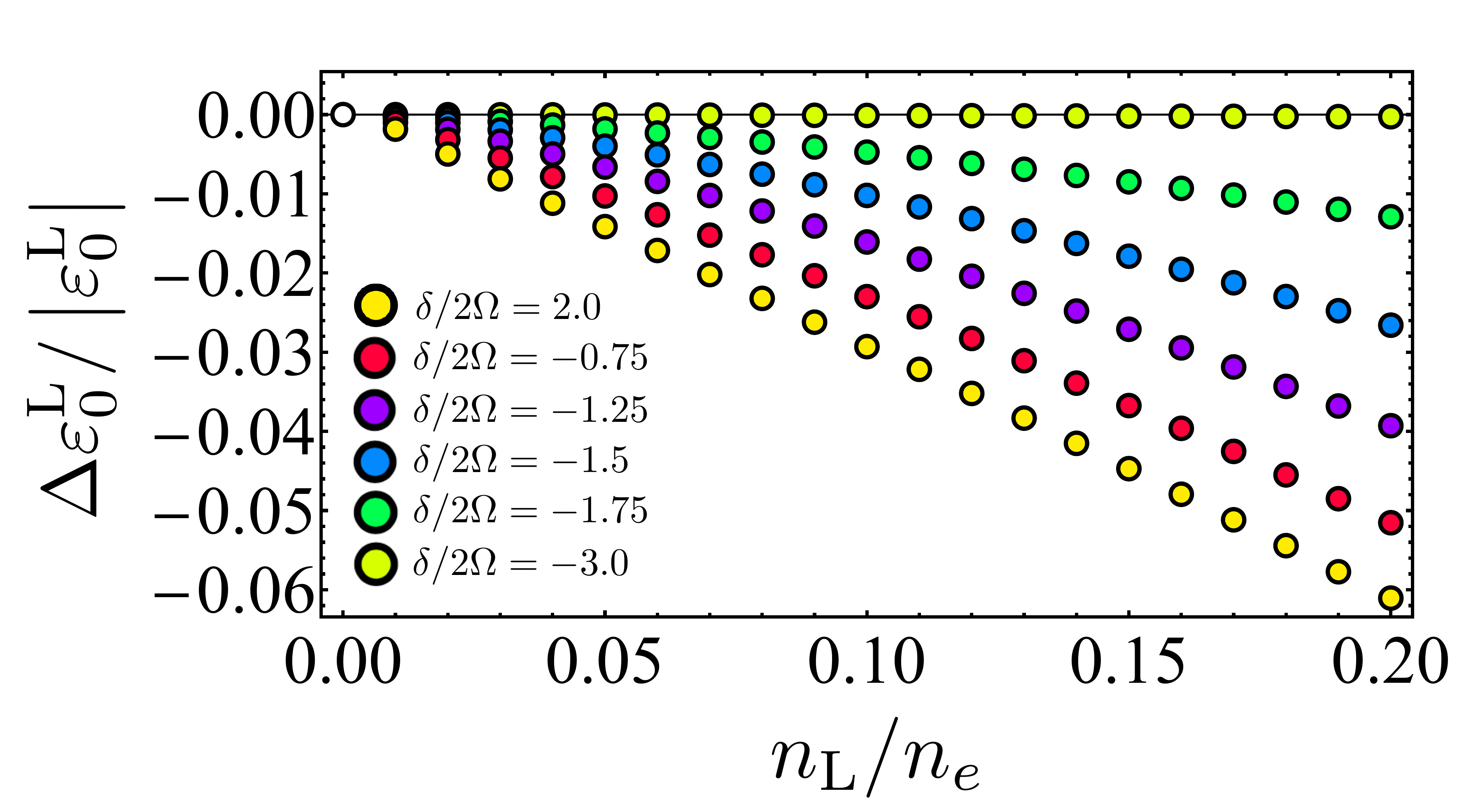} 
\end{tabular}
\end{center}
\vspace{-20pt}
\caption{Energy shift of the $\text{L}$ polaron-polariton branch as a function of their concentration  for representative values of the cavity detuning from $\delta/2\Omega=-3.0$ to $2.0$. The color coding is indicated in the figure. We employ a finite but small temperature $\beta\varepsilon_{F}=0.1$}
\label{fig:4} 
\end{figure}

A non-zero exciton density enters the self-energy explicitly via the second term in Eq.~\eqref{eq:ss}, which comes from the branch-cut of the exciton-electron scattering matrix. Physically, it corresponds to the propagation of an electron and an exciton with population $n_{F}(\omega)$. The exciton density also enters the scattering matrix $\mathcal{T}$ via the exciton-electron pair propagator given by Eq.~\eqref{eq:5}. In Fig.~\ref{fig:4},  we plot the energy shift of the lowest polaron-polariton branch $\Delta\varepsilon_{{\mathbf q}\text{L}}=\varepsilon_{{\mathbf q}\text{L}}-\varepsilon_{{\mathbf q}\text{L}}^{0}$ for ${\mathbf q}=0$ as a function of its density $n_L={\mathcal A}^{-1}\sum_{\mathbf q}n_B(\xi_{{\mathbf q}L})$ for several values of the cavity detuning. Here, $\varepsilon_{{\mathbf q}\text{L}}^{0}$ denotes the energy of the lower polaron-polariton branch in the limit of vanishing density consistent with the notation in section \ref{sec:Landau}. The energy shift is obtained by solving Eq.~\eqref{eq:poles} for a varying chemical potential of the polaritons. We see that the energy shift is \emph{negative} and depends approximately linearly with density $n_L$. From Landau theory, this negative shift corresponds to an \emph{attractive interaction} between the quasiparticles as can be seen explicitly from Eq.~\eqref{eq:LandauQPenergy}.
 
To derive a microscopic expression for the interaction between the polaron-polaritons, it follows from Eq.\ \eqref{eq:LandauInt} that we must evaluate the derivative of the exciton self-energy with respect to their distribution $n_{{\mathbf{q}}\sigma}=n_B(\xi_{{\mathbf q}\sigma})$. We thus expand Eq.~\eqref{eq:ss} as $\Sigma_{xx}(\mathbf{k},\omega)=\Sigma_{n_{\sigma}=0}(\mathbf{k},\omega)+\delta\Sigma(\mathbf{k},\omega)+\mathcal{O}(n_\sigma^{2}),
$
and evaluating this on-shell with $\omega=\xi_{\mathbf{k}\sigma}$ one obtains~\cite{Bastarrachea2021}
\begin{align} \label{eq:4d4}
\frac{\partial\Sigma_{xx}(\mathbf{k},\xi_{\mathbf{k}\sigma})}{\partial n_{{\mathbf{k}}'\sigma'}}=&
\mathcal{X}_{\mathbf{\mathbf{k}'}\sigma'}^{2}\int\!\frac{d^{2}\mathbf{p}}{(2\pi)^{2}}\frac{1}{\xi_{\mathbf{k}\sigma}-\xi_{\mathbf{k}'\sigma'}+\xi_{\mathbf{p}e}-\xi_{\mathbf{k}-\mathbf{k}'+\mathbf{p}e}}\times \nonumber \\ 
&\left[n_{F}(\xi^e_{\mathbf{p}})\mathcal{T}^{2}(\mathbf{k}'-\mathbf{p},\xi_{\mathbf{k}\sigma}+\xi_{\mathbf{p}}^{e})-n_{F}(\xi^{e}_{\mathbf{k}-\mathbf{k}'+\mathbf{p}})
\mathcal{T}^{2}(\mathbf{k}'-\mathbf{p},\xi_{\mathbf{k}-\mathbf{k}'+\mathbf{p}}^{e}+\xi_{\mathbf{k'}\sigma'})\right].
\end{align}
Here it is understood that all energies $\xi_{\mathbf{k}\sigma}$ as well as the $\mathcal{T}$ matrix are evaluated for vanishing quasiparticle density. This 
expression can be generalised to a non-zero density by using  the full density-dependent $\mathcal{T}$-matrix as  shown in  Appendix A. Note that since we are using a non self-consistent approximation, it is the density of the bare upper and polaritons that enter inside the exciton self-energy. To derive Eq.~\eqref{eq:4d4}, we have identified these densities with those of the polaron-polaritons, which corresponds to the first step in a self-consistent calculation. 

\begin{figure}[!ht]
\begin{center}
\begin{tabular}{c}
\includegraphics[width=0.5\columnwidth]{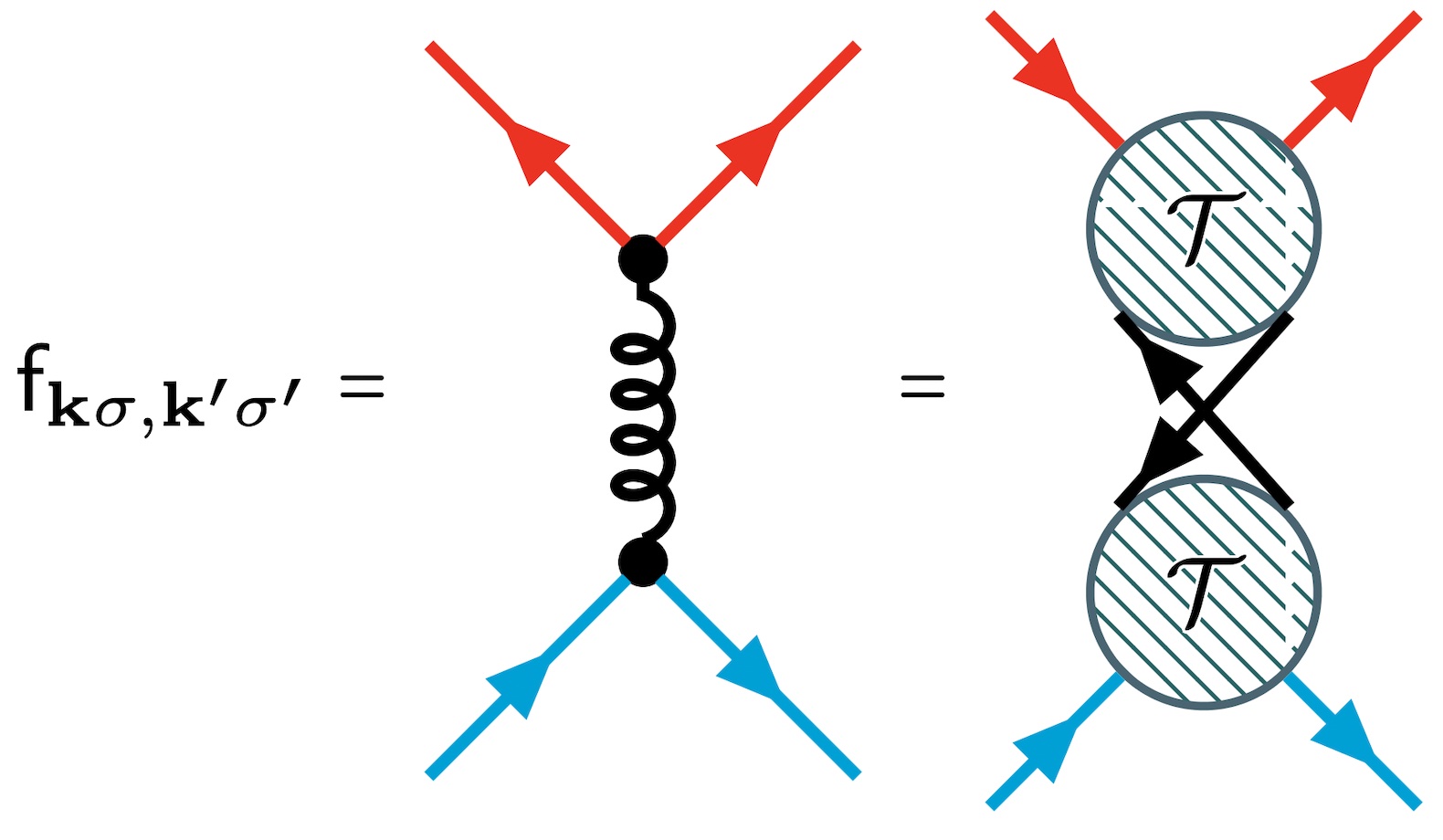} 
\end{tabular}
\end{center}
\vspace{-20pt}
\caption{Feynman diagram of the interaction between quasiparticles $\sigma$ (red lines) and $\sigma'$ (cyan lines) mediated by the 2DEG. The wiggly line corresponds to the induced interaction which translates to a $\mathcal{T}$-matrix repeated scattering mediated by an electron-hole pair (black lines) in the 2DEG.}
\label{fig:3} 
\end{figure}

The effective interaction between polaron-polaritons in branches $\sigma$ and $\sigma'$ 
with momenta ${\mathbf k}$ and  ${\mathbf k}'$ can now be obtained by inserting Eq.\ \eqref{eq:4d4} in Eq.\ \eqref{eq:LandauInt}. Equation \eqref{eq:4d4} is illustrated diagrammatically in Fig.~\ref{fig:3}, which shows that it  corresponds to an induced interaction between two polaron-polaritons mediated by particle-hole excitations of the electron gas. Indeed, when the polaron-polariton  energy is detuned far from the trion energy one can approximate the scattering matrices in Eq.~\eqref{eq:4d4} by the constant 
$\mathcal{T}\simeq \mathcal{T}(\mathbf{0},\xi_{\mathbf{k}\sigma})$, and the  interaction becomes proportional to the 2D Lindhard function~\cite{Bastarrachea2021}, which is characteristic of a particle-hole mediated interaction~\cite{Giuliani2005}.  For stronger interaction between the excitons and the electrons, one must retain the full energy and momentum dependence of  the scattering matrix in Eq.~\eqref{eq:4d4}.
 
We now return to Fig.~\ref{fig:4} where the energy shift of the lowest polaritonic branch ($\sigma'=\text{L}$) is shown as a function of the same lowest polariton concentration ($\sigma=\text{L}$). We can understand it in terms of the effective interaction between the lowest polaron-polaritons. The interaction is attractive since the energy shift is negative, and it increases in strength with the detuning $\delta$. The reason for this is two-fold. First, it is the excitonic component that interacts with the electrons and this component increases with the detuning for the lowest polaron-polariton. Second, the energy of the lowest polaron-polariton approaches the trion energy with increasing $\delta$, which gives rise to strong resonant effects in the electron-exciton scattering. As a result, we see from Fig.~\ref{fig:4} that there can be a sizeable negative energy shift of the polaron-polariton due the attractive interaction mediated by particle-hole excitations in the 2DEG. So far, one has instead observed a temporary positive energy shift corresponding to a repulsive interaction, which has been attributed to a non-equilibrium phase filling effect~\cite{Tan2020}. It would thus be very interesting to investigate this further experimentally as the effective interaction between quasiparticles is a key component of Landau's quasiparticle theory and because it may give rise to strong non-linear optical effects~\cite{Bastarrachea2021,Camacho-Guardian2020}. 

\section{Conclusions}
\label{sec:5}
We presented a  theoretical  framework for describing  polaron-polaritons in 2D semiconductors inside optical microcavities.  Microscopic expressions for the parameters 
entering a Landau quasiparticle theory were given, which  
provides a  simple yet accurate way to describe this new system of interacting hybrid light-matter quasiparticles. Our framework is general apart from assuming that the concentration of the quasiparticles is much smaller than the surrounding 
electron gas and that equilibrium theory can be applied. To illustrate the results,  the ladder approximation was then used to explore the system.   
We also proposed a new non-demolition scheme to probe the energy and  residue of the polaron-polaritons via the  Rabi splittings in the light transmission/reflection  spectrum. Finally, we showed that 
the Landau effective interaction between the polaron-polaritons mediated by particle-hole excitations in the electron gas, is attractive.

Our theoretical framework provides a systematic way to analyse current experiments exploring exciton-polaritons in monolayer TMDs~\cite{Sidler2016,Tan2020}. It 
can moreover be extended to study a new class of exciton-polaritons in van der Waals heterostructures with interlayer Feshbach resonances~\cite{Schwartz2021,Kuhlenkamp2021}, hybridised inter- and inter-layer excitons~\cite{Alexeev2019}, dipolaritons~\cite{Togan2018}, and spatially localised excitons~\cite{Zhang2021,Camacho2021}. The rich  features predicted in these systems~\cite{Shimazako2020,Kennes2021} open the door to using polaritons as quantum probes in strongly correlated electronic states~\cite{Shimazaki2021}, and to realise and control strongly interacting photons. An exciting perspective is to explore the regime of higher polaron-polariton concentrations, where many intriguing 
 phases such as a Bose-Einstein condensate of polaron-polaritons~\cite{Julku2021}, superconductivity, and supersolidity~\cite{Cotlet2016} have been predicted.

\appendix
\unskip

\section{Strong coupling polariton interactions}
\label{app:1}

We take the self-energy as calculated in Eq.~\ref{eq:ss}, but without considering the $\mathcal{T}$-matrix real pole, 
\begin{gather} \label{eq:ssa}
\Sigma_{xx}(\mathbf{k},i\omega_{\nu})= \int\frac{d^{2}\mathbf{q}}{(2\pi)^{2}}n_{F}(\xi_{e\mathbf{q}})\mathcal{T}(\mathbf{k}+\mathbf{q},i\omega_{\nu}+\xi_{e\mathbf{q}})
 \\ \nonumber 
+\int\frac{d^{2}\mathbf{q}}{(2\pi)^{2}}\int_{-\infty}^{\infty}\,\frac{d\omega'}{\pi}\,\frac{n_{F}(\omega')\mbox{Im}\mathcal{T}(\mathbf{k}+\mathbf{q},\omega'+i0^{+})}{i\omega_{\nu}-\omega'+\xi_{e\mathbf{q}}},
\end{gather}

Next, we employ the following relationships
\begin{gather}\label{eq:rel}
\mbox{Im}\mathcal{T}=\left[\left(\mbox{Re}\mathcal{T}\right)^{2}+\left(\mbox{Im}\mathcal{T}\right)^{2}\right]\,\mbox{Im}\Pi
\left[\left(\mathcal{T}-i\mbox{Im}\mathcal{T}\right)^{2}+\left(\mbox{Im}\mathcal{T}\right)^{2}\right]\,\mbox{Im}\Pi=\\ \nonumber
\left[\mathcal{T}^{2}-2i\mathcal{T}\mbox{Im}\mathcal{T}-\left(\mbox{Im}\mathcal{T}\right)^{2}+\left(\mbox{Im}\mathcal{T}\right)^{2}\right]\,\mbox{Im}\Pi=
\left[\mathcal{T}^{2}-2i\mathcal{T}\mbox{Im}\mathcal{T}\right]\,\mbox{Im}\Pi.
\end{gather}
This becomes a series over the imaginary part of the pair propagator. We separate the principal and imaginary parts of the pair propagator in Eq.~\ref{eq:5} as
\begin{gather}
\Pi(\mathbf{q},\omega)
=\sum_{\sigma}\int\frac{d^{2}\mathbf{p}}{(2\pi)^{2}}\frac{}{}\mathcal{X}^{2}_{\sigma\mathbf{q}+\mathbf{p}}\left[1-n_{F}(\xi_{e-\mathbf{p}})+n_{B}(\xi_{\sigma\mathbf{q}+\mathbf{p}})\right]\mbox{x}
\\ \nonumber
\left[\mathcal{P}\frac{1}{\omega-\xi_{e-\mathbf{p}}-\xi_{\sigma\mathbf{q}+\mathbf{p}}}-i\pi\delta(\omega-\xi_{e-\mathbf{p}}-\xi_{\sigma\mathbf{q}+\mathbf{p}})\right],
\end{gather}
inserting it in Eq.~\ref{eq:rel} we obtain
\begin{gather}
\mbox{Im}\mathcal{T}(\mathbf{q},\omega)= 
-\pi\left[\mathcal{T}^{2}-2i\mathcal{T}\mbox{Im}\mathcal{T}\right]\mbox{x}\\ \nonumber 
\sum_{\sigma}\int\frac{d^{2}\mathbf{p}}{(2\pi)^{2}}
\mathcal{X}^{2}_{\sigma\mathbf{q}+\mathbf{p}}
\left[1-n_{F}(\xi_{e-\mathbf{p}})+n_{B}(\xi_{\sigma\mathbf{q}+\mathbf{p}})\right]\delta(\omega-\xi_{e-\mathbf{p}}-\xi_{\sigma\mathbf{q}+\mathbf{p}}).
\end{gather}
Substituting this result in the second term of Eq.~\ref{eq:ssa} and using that $n_{F}(x+y)(1-n_{F}(x)+n_{B}(y))=n_{F}(x)n_{B}(y)$, therefore the self-energy reads
\begin{gather} 
\Sigma_{xx}(\mathbf{k},\omega)
=\int\frac{d^{2}\mathbf{q}}{(2\pi)^{2}}\left\{\frac{}{}n_{F}(\xi_{e\mathbf{q}})\mathcal{T}(\mathbf{k}+\mathbf{q},\omega+\xi_{e\mathbf{q}})\right.
\\ \nonumber
\left.-\sum_{\sigma}\int\frac{d^{2}\mathbf{p}}{(2\pi)^{2}}\frac{\mathcal{X}^{2}_{\sigma\mathbf{k}+\mathbf{q}+\mathbf{p}}n_{B}(\xi_{\sigma\mathbf{k}+\mathbf{q}+\mathbf{p}})n_{F}(\xi_{e-\mathbf{p}})}{\omega-\xi_{\sigma\mathbf{k}+\mathbf{q}+\mathbf{p}}+\xi_{e\mathbf{q}}-\xi_{e-\mathbf{p}}+i0^{+}}\right.\mbox{x}
\\ \nonumber
\left.\left[\mathcal{T}^{2}(\mathbf{k}+\mathbf{q},\xi_{e-\mathbf{p}}+\xi_{\sigma\mathbf{k}+\mathbf{q}+\mathbf{p}}+i0^{+})\right.\left.-2i\mathcal{T}(\mathbf{k}+\mathbf{q},\xi_{e-\mathbf{p}}+\xi_{\sigma\mathbf{k}+\mathbf{q}+\mathbf{p}}+i0^{+})\mbox{Im}\mathcal{T}(\mathbf{k}+\mathbf{q},\xi_{e-\mathbf{p}}+\xi_{\sigma\mathbf{k}+\mathbf{q}+\mathbf{p}}+i0^{+})\right]\right\}.
\end{gather}
As explained in the main text, the quasiparticle interactions are given by the functional derivative of Eqs.~\ref{eq:LandauInt} with respect to the quasiparticle distribution~\cite{Baym1991,Camacho2018}
\begin{gather} \label{eq:chi}
\frac{\mathsf f_{\sigma\mathbf{k},\sigma'\mathbf{k}'}}{\mathcal A}
=
Z_{\sigma'\mathbf{k}'}
\frac{\partial\xi_{\sigma\mathbf{k}}}{\partial n_{\sigma'{\mathbf{k}'}}}
=
Z_{\sigma'\mathbf{q}'}\mathcal{X}_{\sigma'\mathbf{k}'}^{2}\frac{\partial\Sigma(\mathbf{k},\xi_{\sigma\mathbf{k}})}{\partial n_{\sigma'{\mathbf{k}'}}},
\end{gather}
this entails the calculation of the derivative of the second part of the self-energy 
\begin{gather}
\frac{\delta}{\delta n_{B}(\xi_{\sigma'\mathbf{k}'})}\left[n_{B}(\xi_{\sigma\mathbf{k}})\left(\mathcal{T}^{2}(\mathbf{k},\omega)-2i\mathcal{T}(\mathbf{k},\omega)\mbox{Im}\mathcal{T}(\mathbf{k},\omega)\right)\right]= \\ \nonumber
\frac{\delta n_{B}(\xi_{\sigma\mathbf{k}})}{\delta n_{B}(\xi_{\sigma'\mathbf{k}'})}\left\{\mathcal{T}^{2}(\mathbf{k},\omega)-2i\mathcal{T}(\mathbf{k},\omega)\mbox{Im}\mathcal{T}(\mathbf{k},\omega)+\right.\\ \nonumber
\left. n_{B}(\xi_{\sigma\mathbf{k}})\left[2\mathcal{T}(\mathbf{k},\omega)-2i\mbox{Im}\mathcal{T}(\mathbf{k},\omega)\right]\frac{\partial \mathcal{T}(\mathbf{k},\omega)}{\delta n_{B}(\xi_{\sigma'\mathbf{k}'})}-n_{B}(\xi_{\sigma\mathbf{k}})\mathcal{T}(\mathbf{k},\omega)
\left(\frac{\partial \mathcal{T}(\mathbf{k},\omega)}{\delta n_{B}(\xi_{\sigma'\mathbf{k}'})}-\frac{\partial \mathcal{T}^{*}(\mathbf{k},\omega)}{\delta n_{B}(\xi_{\sigma'\mathbf{k}'})}
\right)
\right\}.
\end{gather}
The functional derivative of the $\mathcal{T}$-matrix is given by
\begin{gather}
\frac{\delta}{\delta n_{B}(\xi_{\sigma'\mathbf{k}'})}\mathcal{T}(\mathbf{k},\omega)=
\frac{\mathcal{T}_{0}^{2}}{\left[1-\mathcal{T}_{0}\Pi(\mathbf{k},\omega)\right]^{2}}\frac{\delta \Pi(\mathbf{k},\omega)}{\delta n_{B}(\xi_{\sigma'\mathbf{k}'})}=\\ \nonumber
\mathcal{T}^{2}(\mathbf{k},\omega)\sum_{\sigma}\int\frac{d^{2}\mathbf{p}}{(2\pi)^{2}}\frac{\mathcal{X}_{\sigma\mathbf{k}+\mathbf{p}}^{2}}{\omega-\xi_{e-\mathbf{p}}-\xi_{\sigma\mathbf{k}+\mathbf{p}}}\frac{\delta n_{B}(\xi_{\sigma\mathbf{k}+\mathbf{p}})}{\delta n_{B}(\xi_{\sigma'\mathbf{k}'})}=\\ \nonumber
=\mathcal{T}^{2}(\mathbf{k},\omega)\int\frac{d^{2}\mathbf{p}}{(2\pi)^{2}}\mathcal{X}_{\sigma\mathbf{k}+\mathbf{p}}^{2}\frac{\delta_{\sigma,\sigma'}\delta(\mathbf{k}'-(\mathbf{k}+\mathbf{p}))}{\omega-\xi_{e-\mathbf{p}}-\xi_{\sigma\mathbf{k}+\mathbf{p}}+i0^{+}}=
\frac{\mathcal{X}_{\sigma'\mathbf{k}'}^{2}\mathcal{T}^{2}(\mathbf{k},\omega)}{\omega-\xi_{e\mathbf{k}-\mathbf{k}'}-\xi_{\sigma\mathbf{k}'}+i0^{+}}.
\end{gather}
Because the derivative is of the order $\mathcal{T}^{2}$, if we keep only terms associated to second-order diagrammatic contributions we can approximate 
\begin{gather}
\frac{\delta}{\delta n_{B}(\xi_{\sigma'\mathbf{k}'})}\left[n_{B}(\xi_{\sigma\mathbf{k}})\left(\mathcal{T}^{2}(\mathbf{k},\omega)-2i\mathcal{T}(\mathbf{k},\omega)\mbox{Im}\mathcal{T}(\mathbf{k},\omega)\right)\right]\simeq \mathcal{T}^{2}(\mathbf{k},\omega)\delta(\mathbf{k}-\mathbf{k}')\delta_{\sigma,\sigma'}.
\end{gather} 
In this way, after substituting the derivative of the $\mathcal{T}$-matrix into the derivative of the self-energy, the mediated potential on-shell, up to second order diagrams, reads
\begin{gather} 
\frac{\partial\Sigma(\mathbf{k},\xi_{\sigma\mathbf{k}})}{\partial n_{\sigma'{\mathbf{k}'}}}=
%
%
\mathcal{X}_{\sigma'\mathbf{k}'}^{2}\int\frac{d^{2}\mathbf{q}}{(2\pi)^{2}}\frac{1}{\xi_{\sigma\mathbf{k}}-\xi_{\sigma'\mathbf{k}'}+\xi_{e\mathbf{q}}-\xi_{e\mathbf{k}-\mathbf{k}'+\mathbf{q}}+i0^{+}}\mbox{x}
\\ \nonumber
\left[n_{F}(\xi_{e\mathbf{q}})\mathcal{T}^{2}(\mathbf{k}'-\mathbf{q},\xi_{\sigma\mathbf{k}}+\xi_{e\mathbf{q}}+i0^{+})-n_{F}(\xi_{e\mathbf{k}-\mathbf{k}'+\mathbf{q}})\mathcal{T}^{2}(\mathbf{k}'-\mathbf{q},\xi_{e\mathbf{k}-\mathbf{k}'+\mathbf{q}}+\xi_{\sigma'\mathbf{k}'}+i0^{+})\right],
\end{gather}
which is identical to Eq.~\ref{eq:4d4} from the main text.
\bibliography{references}

\end{document}